\documentclass[a4paper]{article}
\usepackage{array,booktabs,tabularx,threeparttable}
\usepackage{float,placeins}
\usepackage{arxiv}
\usepackage[square,numbers,sort&compress]{natbib}
\usepackage[utf8]{inputenc} 
\usepackage[T1]{fontenc}    
\usepackage{hyperref}       
\usepackage{url}            
\usepackage{booktabs}       
\usepackage{amsfonts}       
\usepackage{nicefrac}  
\usepackage{adjustbox}
\usepackage{microtype}      
\usepackage{lipsum}
\usepackage{xspace}
\usepackage{lipsum,amsmath,amssymb,amsfonts}
\usepackage{graphicx}
\graphicspath{ {./images/} }

\title{Physics-Informed Neural Networks for Industrial Gas Turbines: Recent Trends, Advancements and Challenges}

\author{
 Afila Ajithkumar Sophiya \\
  Lincoln AI Lab\\
  School of Engineering and Physical Science\\
  University of Lincoln\\
  Lincoln, United Kingdom\\
  \texttt{29511866@students.lincoln.ac.uk} \\
  \And
 Sepehr Maleki \\
   Lincoln AI Lab\\
  School of Engineering and Physical Science\\
  University of Lincoln\\
  Lincoln, United Kingdom\\
  \texttt{smaleki@lincoln.ac.uk} \\
   \And
 Giuseppe Bruni \\
  Siemens Energy\\
  Lincoln, United Kingdom\\
  \texttt{} \\
   \And
 Senthil K. Krishnababu \\
  Siemens Energy\\
  Lincoln, United Kingdom\\
  \texttt{} \\
}
\date{}

\begin{document}
\maketitle
\begin{abstract}
Physics-Informed Neural Networks (PINNs) have emerged as a promising computational framework for solving differential equations by integrating deep learning with physical constraints.
However, their application in gas turbines is still in its early stages, requiring further refinement and standardization for wider adoption. This survey provides a comprehensive review of PINNs in Industrial Gas Turbines (IGTs) research, highlighting their contributions to the analysis of
aerodynamic and aeromechanical phenomena, as well as their applications in flow field reconstruction,
fatigue evaluation, and flutter prediction, and reviews recent advancements in accuracy,
computational efficiency, and hybrid modeling strategies. In addition, it explores key research
efforts, implementation challenges, and future directions aimed at improving the robustness and
scalability of PINNs.
\end{abstract}
\keywords{Physics-informed neural networks
(PINNs), Industrial gas turbines (IGTs)}


\section{Introduction}
\noindent Industrial Gas Turbines (IGTs) are the cornerstone of modern energy and transportation systems, providing efficient and reliable power generation and propulsion. Their performance, reliability and durability are highly dependent on aerodynamic and aeromechanic principles~\cite{dixon2013fluid, Cumpsty_2003}. However, turbines operating in extremely challenging conditions, such as high pressure, temperature, and velocity, face many concerns. These include flow instability, which can cause compressor surge or stall, turbulence that complicates accurate modeling and reduces aerodynamic efficiency, and blade flutter and forced vibrations, which can lead to fatigue and structural failure of rotating components. In addition, steep temperature gradients can result in thermal stresses that cause materials to creep, oxidize, and degrade their coating. These can hinder the performance and shorten the lifespan of components, but also increase fuel emissions when not properly analyzed~\cite{lefebvre2010gas,dixon2013fluid, Cumpsty_2003}. Although analytical methods have been widely used as the foundation for gas turbine design and analysis, especially in the early stages, high-fidelity simulation techniques, such as Finite Element Method (FEM) and Computational Fluid Dynamics (CFD)\cite{IACOVIDES1995454, 10.1115/2001-GT-0267, Deepanraj_Lawrence_Sankaranarayanan_2011}, have been used to resolve detailed flow and structural behavior. The computational and time-consuming nature of these simulations, especially for unsteady, 3D, and multiphysics problems, restricts their use in real-time or large-scale design optimization~\cite{tucker2013unsteady}.

\medskip
\noindent Machine Learning (ML) and data-driven approaches have been used to analyze and predict these complex behaviors in gas turbines~\cite{PACHAURI2024131421}. Although these approaches offer a faster and more scalable alternative~\cite{10.1115/1.4055634}, they often lack physical consistency, leading to inaccuracies when extrapolated beyond the training data~\cite{osti_2282016}. To address these limitations, Physics-Informed Neural Networks (PINNs)~\cite{RAISSI2019686} have emerged as a promising solution. The key idea of PINNs involves the integration of physical laws, represented as Partial Differential Equations (PDEs), directly into the neural network loss function. For instance, PINNs incorporate fundamental governing equations like the Navier-Stokes equations (N-S) into network's loss to accurately model fluid flow phenomena. This approach ensures that predictions not only rely on available data, but also adhere to fundamental principles such as the conservation of mass, momentum, and energy. This embedding of physics constraints can effectively provide accurate predictions even in scenarios with sparse data, significantly enhancing their applicability to address complex gas turbine problems.

\medskip
\noindent PINNs have been extensively adopted across diverse fields~\cite{AMIN2025116728,ZHANG2025213689,AGHAEE2025109528} due to their advantages over purely data-driven models. By integrating physical laws directly into the network structure, PINNs reduce their dependence on large datasets and enhance their predictive capabilities for unseen operating conditions~\cite{RAISSI2019686,osti_2282016}. This approach offers substantial improvements in simulation accuracy, reduces computational costs, and supports real-time decision making in IGTs operations. The incorporation of PINNs has great potential to transform the analysis of aerodynamic and aeromechanical challenges, ultimately driving advances in turbine efficiency, reliability, and sustainability. This study focuses on comprehensively presenting the existing literature on PINNs to solve IGTs problems, highlighting their diverse applications in addressing aerodynamic and aeromechanical challenges. 
\section{Contributions}
\noindent Several comprehensive surveys provide valuable insights into the general theory~(\cite{cuomo2022scientific,ai5030074,bdcc6040140,sophiya2025comprehensiveanalysispinnsvariants}) and applications~(\cite{osti_2282982, 9743327, 10.1063/5.0226562}) of PINNs. However, there remains scope for a more detailed exploration of PINNs tailored specifically to IGTs applications.

\noindent The contribution of this survey can be summarized as follows:
\begin{enumerate}
    \item \textbf{Overview of PINNs methodology and their theoretical foundation}: A brief study on the fundamental principles of PINNs and their ability to incorporate physical laws into deep learning models, providing a theoretical background on loss function formulation, training strategies, and convergence properties, thereby establishing a foundation for their application in industrial gas turbines.
      \item \textbf{Recent advancements in PINNs}: A detailed review of cutting-edge techniques and recent developments that have significantly enhanced accuracy and computational efficiency.
    \item \textbf{Limitations and future scope}: The study discusses current limitations and explores potential improvements for future advancements in gas turbine research.
    \item \textbf{Application of PINNs in aerodynamics and aeromechanics}: The study explores how PINNs are increasingly being used to model aerodynamic flow behavior and structural mechanics in gas turbines. This includes use cases like flow field reconstruction, fatigue analysis and flutter prediction, offering data-efficient alternatives to CFD methods.
\end{enumerate}
The remainder of this paper is divided into eight sections. Section \ref{sec2} provides a brief analysis of various deep learning architectures used to solve IGTs related problems. Section \ref{sec3} introduces the foundational PINNs architecture and key concepts. Sections \ref{sec4} and \ref{sec5} explore the application of PINNs within aerodynamics and aeromechanics, respectively. Section \ref{sec6} highlights recent advancements of PINNs in these domains. Section \ref{sec7} addresses common implementation challenges, and finally, Section \ref{sec8} provides concluding remarks.
\section{Machine Learning in Industrial Gas Turbines}\label{sec2}
\noindent ML surrogate models have gained substantial interest in modeling and analyzing IGTs, where high-dimensional, nonlinear, and multi-physics problems challenge traditional numerical methods. Over the past few years, these models have been used to tackle various IGTS tasks. Condition assessment, fault detection, and health monitoring~(\cite{en14248468}) of turbines are a few of these tasks. There is also a significant impact from these ML models~(\cite{zou2024application}) in aerodynamic analysis and modeling of different components of the turbomachinery.

\medskip 

\noindent Several prominent Deep Learning (DL) architectures have been explored within the context of gas turbine research  particularly for tasks like flow prediction, structural analysis, and system diagnostics. Among these, Convolutional Neural Networks (CNNs) have shown significant promise as surrogate models. Their ability to approximate high-dimensional outputs such as flow fields, pressure distributions, structural deformations, and stress concentrations has made them a viable alternative to computationally intensive simulations. Recent studies~(\cite{17437169320240101,10522978,10.1063/5.0186087,Bruni_Maleki_Krishnababu_2025}) highlight CNNs is a promising technique for predicting the flow field and aerodynamics of turbo machinery components. However, due to the requirement for structured grids, CNNs are limited to the 3D domain. Graph neural networks~(\cite{harsch2021directpredictionsteadystateflow}) helped to overcome this, however, due to the dependence of CFD grids, issues with scalability and flexibility still exist, restricting their use in real-world scenarios.

\medskip

\noindent In addition to spatial modeling, Recurrent Neural Networks (RNNs) and Long Short-Term Memory (LSTM) networks~(\cite{Shuai_2023,7385755}) are well-suited for modeling time-dependent behaviors, making them particularly valuable in gas turbine applications involving dynamic phenomena, forecasting performance degradation, and modeling unsteady flow behaviors. However, RNNs struggle with long-term dependencies~(\cite{10.1115/GTINDIA2023-118326}) and high-dimensional spatial representations, which are critical for accurately modeling unsteady flows.
\medskip

\noindent Another class of models, Generative Adversarial Networks (GANs) have been explored for turbine blade design and fault diagnostics, particularly when data is sparse~(\cite{WANG2022123373,Yao_2023}). Their ability to generate high-fidelity output makes them appealing to provide fast aerodynamic predictions. However, ensuring training stability and preventing mode collapse~(\cite{barsha2025depth}) remains active challenges. Studies are exploring physics-informed GANs~(\cite{wang2023physics,li2022using}) variants that effectively simulate complex systems such as fluid flows and structural mechanics.

\medskip

\noindent Autoencoders are employed for anomaly detection and feature extraction in IGTs~(\cite{9187054,10.1115/GT2024-129046}). However, due to the black box nature, the predicted outcomes might not be interpretable and reliable.  To tackle this issue, researchers have investigated combining output consolidation and autoencoders to increase the accuracy of performance prediction~(\cite{pongetti2021using}). However, there are still issues with generalizability in these models.
\medskip

\noindent Although DL techniques have advanced modeling capabilities in IGTs applications, several inherent limitations restrict their broader applicability in scientific and engineering domains. Most notably, they lack embedded physical constraints, making them prone to errors when extrapolating beyond the training data. Additionally, the limited availability of labeled data, which is caused by its high cost and impracticability to obtain in real-world settings, further hinders their effectiveness. These challenges have motivated the development of hybrid approaches that integrate data-driven learning with physical principles. Among them, PINNs have emerged as a promising framework that combines the strengths of deep learning with the rigor of physics-based modeling.
\section{Overview of PINNs}\label{sec3}
\noindent Utilizing DL to solve PDEs~(\cite{712178}) dates back to the late 1990s. However, it was only in 2019 that \cite{RAISSI2019686} coined the term PINNs, combining deep neural networks to effectively solve various engineering problems modeled by PDEs, incorporating prior knowledge and physical constraints into the architecture of neural network. A key strength of PINNs lies in their versatility that they can directly model physical phenomena or infer missing parameters.
\medskip

\noindent Consider the incompressible N-S equation in two dimension:
\begin{equation}
\begin{aligned}
    \frac{\partial u}{\partial x}+\frac{\partial v}{\partial y}&=0\\[6pt]
    \frac{\partial u}{\partial t}+u\frac{\partial u}{\partial x}+v\frac{\partial u}{\partial y}&=-\frac{1}{\rho}\frac{\partial p}{\partial x}+\mu\left(\frac{\partial^2 u}{\partial x^2}+\frac{\partial^2 u}{\partial y^2}\right)\\[6pt]
    \frac{\partial v}{\partial t}+u\frac{\partial v}{\partial x}+v\frac{\partial v}{\partial y}&=-\frac{1}{\rho}\frac{\partial p}{\partial y}+\mu\left(\frac{\partial^2 v}{\partial x^2}+\frac{\partial^2 v}{\partial y^2}\right)
\end{aligned}
\end{equation}

\noindent where $t$ is the time variable, $(x,y)\in \Omega \subset \mathbb{R}^2$, $p$ represents the pressure field and $u$ and $v$ represent the velocity components in the $x$ and $y$ directions, respectively. The fluid density is represented by $\rho$ and the kinematic viscosity of the fluid by $\mu$.
\medskip

\noindent The same equation can be represented in a more simplified form and can be re-stated as follows:
\begin{equation}
\begin{aligned}
 f(\textbf{x}, \textbf{u}, p, \partial_\textbf{x}\textbf{u}, \partial_\textbf{x}p, \cdots, \omega)&= 0,  \textbf{x}\in \Omega\\
 \textbf{u}(\textbf{x})&=\textbf{I}(\textbf{x}), \textbf{x}\in \Omega\\
   \textbf{u}(\textbf{x})&=\textbf{B}(\textbf{x}), \textbf{x}\in \partial\Omega
\end{aligned}
\end{equation}
where $f$ signifies the residual of the N-S equations, incorporating differential operators of velocity $\textbf{u}$ and pressure $p$.  It may also include additional parameters $\omega$ for inverse problem identification.  $\textbf{x}$ specifies the $2D$ coordinates within the computational domain, while $\textbf{I}$ defines the initial condition and $\textbf{B}$ the boundary conditions.
\medskip

\noindent Fundamentally, the PINNs architecture comprises three components: a neural network, physics constraints, and a tailored loss function. The first component, the neural network, acts as a universal approximator~(\cite{HORNIK1989359}). It accepts the input comprising both the prior knowledge obtained from the experimental data and the data sampled from the computational domain. The neural network outputs $\hat{\textbf{u}}_\theta$, which serves as a surrogate or approximate solution to the true function $\textbf{u}$. The term $\theta$ comprises all the tunable parameters, including both weights and biases, which are optimized during training, thus ensuring that the network accurately approximates the underlying physical behavior. In the case of N-S equation the output generally comprises of the velocity components, $\hat{\textbf{u}}_\theta$ and the pressure component, $\hat{p}_\theta$, which are the flow parameters of the equation.  Although fully connected neural networks (FNN) are commonly employed in PINNs due to their simplicity and effectiveness, alternate architectures such as Convolutional Neural Networks (CNNs), Recurrent Neural Networks (RNNs), and Generative Adversarial Networks (GANs)~(\cite{soltani2024tradeoffreconstructionaccuracyphysical,10.1007/978-3-031-08754-7_45,CIFTCI2024116907}), have also been explored in the literature of PINNs.  The second component consists of embedding the governing equations, along with the initial and boundary conditions, into the training process of the network through Automatic Differentiation (AD)~(\cite{JMLR:v18:17-468}). AD computes exact derivatives by systematically applying the chain rule during backpropagation, which ensures high precision and numerical stability. The third component is the specially designed loss function, often referred to as the feedback mechanism. The PINNs loss function is typically structured to incorporate both physics loss and data loss.  The physics-based loss enforces adherence to the embedded governing equations, initial conditions, and boundary conditions. The data loss, on the other hand, incorporates available labeled data, such as measured pressure, temperature, or velocity fields, providing empirical grounding to the model. It is worth noting that not all PINNs require labeled datasets for training, which is considered to be the primary advantage over conventional data-driven models. In contrast to ML approaches, PINNs can be trained on data-sparse regimes by leveraging embedded physical laws. They allow unsupervised or semi-supervised training, which significantly reduces their reliance on large datasets. The loss function for the corresponding equation can be expressed as:
\begin{equation}
    \mathcal{L}(\theta) = \lambda_{PDE} \cdot  \mathcal{L}_{PDE} + \lambda_{data} \cdot  \mathcal{L}_{data} + \lambda_{BC} \cdot  \mathcal{L}_{BC} + \lambda_{IC} \cdot  \mathcal{L}_{IC}
\end{equation}
where $\mathcal{L}_{PDE}$ represents the residual loss of N-S equation, $\mathcal{L}_{data}$ represents the data loss that measures discrepancies between the network predictions and labeled observational data and $\mathcal{L}_{BC}$ and $\mathcal{L}_{IC}$ enforce boundary and initial conditions. Also, the $\lambda$ values represent the weights assigned for each loss term. The weights, which are often considered as hyperparameters, are crucial for balancing the contribution of physics-based and data-based constraints in the loss function. Although manual tuning is still commonly used, studies~(\cite{wang2021understanding}) have proposed adaptive strategies for dynamically adjusting these weights to improve convergence and decrease overfitting. The residual loss of the governing equation can be computed as follows:
\begin{equation}
    \mathcal{L}_{PDE}=\frac{1}{N_{PDE}}\sum_{i=1}^{N_{PDE}}\left |\left|f(\textbf{x}_i, \hat{\textbf{u}}, \hat{p}, \partial_\textbf{x}\hat{\textbf{u}}, \partial_\textbf{x}\hat{p}, \cdots, \omega)\right|\right|^2_2
\end{equation}
The mean-square error (MSE) of the N-S equation which is evaluated with $N_{PDE}$ points sampled from inside the computational domain $\Omega$. Likewise, similar pattern can be adapted to compute the prediction loss $\mathcal{L}_{data}$ with $N_{data}$ number of labeled training points.
\begin{equation}
    \mathcal{L}_{data}=\frac{1}{N_{data}}\sum_{i=1}^{N_{data}}\left |\left| ( \hat{\textbf{u}}_\theta- \textbf{u}_i)+( \hat{p}_\theta- p_i)\right|\right|^2_2
\end{equation}
Similarly, the initial and boundary loss can also be formulated with $N_{IC}$ and $N_{BC}$ points from the domain and boundary respectively.
\begin{align}
   \mathcal{L}_{IC}&= \frac{1}{N_{IC}} \sum_{i=1}^{N_{IC}}\left |\left| \hat{\textbf{u}}(\textbf{x}_i)-\textbf{I}(\textbf{x}_i)  \right|\right|^2_2 \\
   \mathcal{L}_{BC}&=\frac{1}{N_{BC}}\sum_{i=1}^{N_{BC}}\left |\left|  \hat{\textbf{u}}(\textbf{x}_i)-\textbf{B}(\textbf{x}_i)  \right|\right|^2_2
\end{align}

\begin{figure}
    \centering
    \includegraphics[width=1\linewidth]{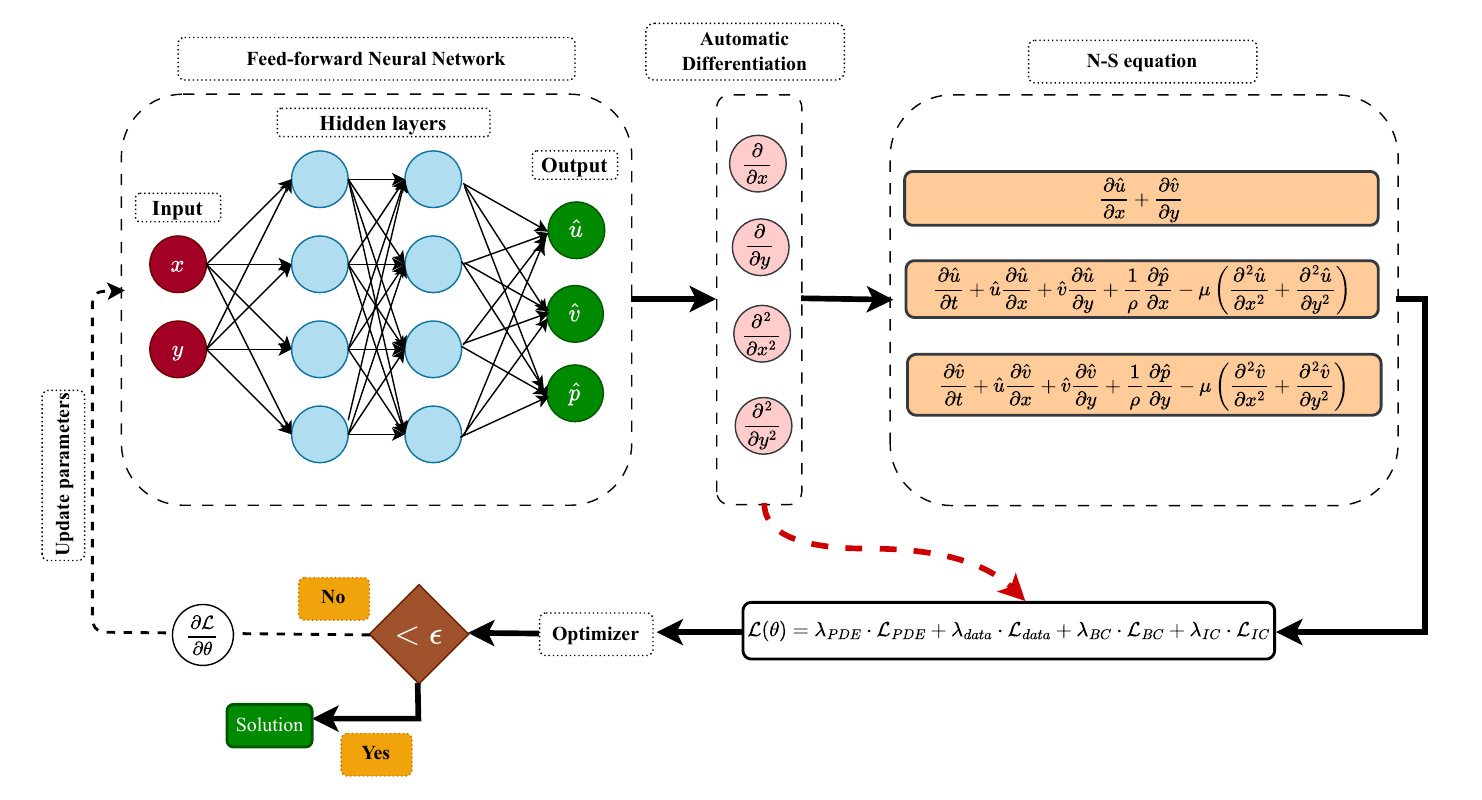}
    \caption{PINNs architecture}
    \label{figure1}
\end{figure}

\noindent Once the loss function is formulated, the model is optimized to achieve better performance. A widely used optimization method for training PINNs includes the Adam optimizer and quasi-Newton methods like L-BFGS (Limited-memory Broyden-Fletcher-Goldfarb-Shanno). Recent practice adopts hybrid optimization strategy~(\cite{doi:10.1137/19M1274067}), where the model is initially trained using Adam to rapidly converge towards a near-optimal solution, followed by L-BFGS for fine-tuning to achieve higher accuracy and smoother convergence. Figure~\ref{figure1} illustrates the basic architecture of a PINNs model, highlighting the arrangement of its various components.
\medskip

\noindent  The architecture and training strategy of PINNs are critical factors that directly influence their effectiveness in solving complex physical problems. Recent advancements have introduced innovative sampling techniques such as adaptive sampling, residual-based adaptive refinement (RAR)~(\cite{doi:10.1137/19M1274067}), and importance sampling~(\cite{https://doi.org/10.1111/mice.12685}) which enhance model accuracy by prioritizing regions with higher residuals or more complex physical behavior. These methods ensure better coverage of the computational domain and lead to faster convergence, especially in scenarios with sharp gradients or localized turbulence. Additionally, the use of adaptive learning rate schedules, including strategies like cosine annealing~(\cite{2023AcMSn..3922302X}), exponential decay, and learning rate warm-up, has proven effective in improving the convergence rate and stability of PINNs during training. These techniques allow for dynamic adjustment of the learning rate, enabling more efficient exploration of the loss landscape and reducing the likelihood of getting trapped in local minima. Various PINNs architectures~(\cite{MENG2020113250,doi:10.1137/18M1229845}) have been proposed, including single-network and segregated-network approaches. Each architecture presents its own set of trade-offs concerning accuracy and computational cost~(\cite{shukla2021parallel}). The optimal choice depends on the specific problem being addressed and the available computational resources.

\section{PINNs for aerodynamics}\label{sec4}
\noindent The aerodynamic performance of gas turbine components, such as compressor, turbine and combustor are governed by a complex interplay of fluid dynamics phenomena, posing significant challenges for accurate modeling and optimization. Conventional approaches(\cite{18116454920241115,17535987220240201,Almensoury_2021}), such as high-fidelity CFD simulations and reduced-order models, are computationally expensive, require significant domain expertise, and may struggle to effectively capture transient and multi-physics behavior. However, DL methods are becoming increasingly appealing for modeling and optimizing turbine performance under operational constraints, as they support real-time prediction, uncertainty quantification, inverse design, and the integration of sparse sensor data. This section categorizes and summarizes key studies that have applied PINNs to various aerodynamic challenges in IGTs. Table \ref{tab1} presents a concise comparison of these works, highlighting their methodologies, key findings, and limitations.

\subsection{Foundational Aerodynamic Modelling with PINNs }

\noindent The early adoption of PINNs in IGTs aerodynamics focused on solving core aerodynamic equations such as the Euler and N-S equations. These foundational models serve as benchmarks to validate the feasibility and robustness of PINNs in fluid mechanics. \cite{MAO2020112789} applied PINNs to model high-speed aerodynamic flows by solving the Euler equations in both forward and inverse settings. The study
demonstrated accurate prediction of flow variables such as velocity and pressure, particularly in regions free from shocks or discontinuities in the steady flow. For problems with contact discontinuities, the clustered points around discontinuities significantly improved predictive accuracy without prior knowledge of their location. While the predictions for velocity and pressure remained highly accurate, the density predictions exhibited higher errors due to the presence of discontinuities. Another paper is by \cite{10.1063/5.0095270}, proposed a PINNs based framework for solving the RANS equations in incompressible turbulent flows using only boundary data, without relying on traditional data-driven models. The study demonstrated that PINNs can accurately capture key aerodynamic features, including mean flow characteristics and Reynolds-stress components across varying flow regimes. However, a noticeable change in accuracy can be observed as the Reynolds number($Re$) increases from a low to a high value, due to the increasing complexity of turbulence scales.
\medskip

\noindent \cite{CAI20251} proposes a  method to determine pressure fields from velocity components without explicitly incorporating boundary conditions in the loss function. The approach enforces the N-S equations as a physical constraint to model incompressible flow, with input velocity features pre-processed using decimal scaling for numerical stability. The  model demonstrated generalizability across multiple flow scenarios, including oblique Hiemenz flow, flow over a NACA0012 airfoil, and a hawkmoth (Manduca) flight experiment. The study also assessed robustness under noisy inputs, confirming the ability of the model to reconstruct pressure fields accurately. \cite{10.1007/978-3-030-59851-8_9} discusses a novel methodology  to determine the optimal parameters in RANS $k-\epsilon$ turbulence models using a physics-embedded neural network approach, leveraging limited availability of high-fidelity DNS data. This approach enhances the predictive capability of RANS models while maintaining adherence to fundamental physical laws, reducing reliance on empirical parameter tuning. The method was validated on a flow over the bump dataset, leveraging high-resolution velocity and pressure field data. This parameter training proved crucial for accurately predicting the evolution of flow regions during adverse pressure gradients. However, the inherent simplifications within RANS models limit their accuracy for high-turbulence flows. \cite{LI2022114524} introduce ReF-nets, a PINNs framework designed to solve the Reynolds equation governing gas bearing lubrication. The prediction accuracy and physics interpretability were investigated through supervised, unsupervised, and semi-supervised learning approaches. While not directly focused on aerodynamics, the study highlights the trade-offs between accuracy and physical interpretability in different learning modes, offering insights that could inform the development of interpretable PINNs models for gas turbine applications.
\subsection{Turbulence Modeling}

\noindent Turbulence modeling remains a significant challenge in CFD, particularly in resolving intricate, unsteady flow structures in compressors and turbines due to high computational costs (\cite{10.1115/1.4064367}). However, PINNs have emerged as a promising approach to address these limitations. \cite{fluids8020043} explored the use of PINNs to predict turbulent flow characteristics over a backward-facing step at $Re=5100$. The PINNs were trained using sparse labeled data, specifically velocity measurements along three to five vertical lines downstream of the step.  The result demonstrated that increasing the number of vertical lines improved the accuracy of flow predictions across all models, highlighting the importance of data density in training PINNs. A similar work was done by \cite{HANRAHAN2023109232} to model time-averaged turbulent flows without relying on traditional turbulence closure models. The study solves open RANS equations using PINNs with sparse experimental data on Adverse Pressure Gradient (APG) boundary layers and periodic hills. Results showed that PINNs maintained predictive accuracy with significant data reduction, emphasizing their potential in data-scarce environments common in gas turbine applications. Additionally, it was observed that the accuracy of Reynolds stress predictions degraded with increasing Reynolds number, particularly for $Re > 20,000$, highlighting ongoing challenges in capturing fine-scale turbulence features.
\medskip

\noindent \cite{LIU2024113275} proposed a PINNs framework for reconstructing three-dimensional turbulent combustion fields including velocity, temperature, and species mass fraction, from sparse and noisy datasets. The model employed a FNN equipped with sinusoidal activation functions and self-adaptive linear layers, integrating both residual and sparse data points as inputs. Synthetic sparse data was generated using DNS for two flame configurations: freely propagating planar premixed flames and slot-jet premixed flames. To relicate experimental data, three categories of sparse datasets were considered: random sampling, regular sampling, and noisy random sampling.
Separate networks were trained for each governing equation (momentum, species mass
fraction, and temperature), and the results showed strong agreement with DNS data, validating the model's capability in handling sparse inputs. 
However, performance was sensitive to network size and degraded in capturing small-scale flame structures at high Reynolds numbers. These findings highlight the potential of PINNs in high fidelity combustion modeling under data limited conditions, with future work needed to address challenges in resolving small-scale turbulence.
\medskip

\noindent \cite{10.1115/GT2023-101238} employed PINNs to predict the translational and turbulent flows around turbine blades, focusing on critical aerodynamic features such as separation bubble length and wake loss profiles aspects that are challenging to model using conventional traditional RANS models. The study investigated two benchmark problems: periodic hills with $Re=10595$ and a low-pressure turbines with $Re=80000$. For the periodic hill case, the PINNs were trained with layer-wise locally adaptive activation functions on the incompressible flow equation, and  accurately predicted velocity, pressure, and Reynolds stress profiles, though boundary errors appeared due to sparse data. For the $T106C$ turbine blade case, PINNs were tested under compressible and incompressible flow models across different turbulence intensities. While PINNs effectively predicted Reynolds stress and turbulent heat flux, errors arose at the leading and trailing edges due to sparse data availability. Additionally, compressible flow modeling proved computationally expensive, requiring a high number of residual points and larger network architectures. The study also highlighted the impact of insufficient data coverage in certain regions, leading to localized inaccuracies. 
\subsection{Airfoil and Blade Design}

\noindent PINNs are increasingly being applied to aerodynamic applications in industrial setups, particularly in airfoil design. \cite{10.1063/5.0188665} proposed a method combining PINNs with mesh transformation, termed NNfoil, to learn fluid flows near the leading edge of an airfoil. The method enables to solve the two-dimensional Euler equations more effectively by converting the complex physical geometry of the aerofoil into a uniform computational domain. This transformation enhances numerical accuracy and stability, especially when it comes to capturing sharp gradients in inviscid subsonic flows around curved surfaces. The neural network output, which utilized volume weighted PDE residuals in the loss function, was compared with the classical finite volume method (FVM) for different flow conditions.
The results obtained exhibited consistent agreement with FVM results and the NNfoil accurately predicted the flow around thinner airfoils. An extended version by \cite{CAO2024113285} was proposed with a complete state-space solution model termed NNfoil-C to solve subsonic inviscid flows around airfoils. This model incorporates parameters that encompass a broad spectrum of flow conditions and airfoil shapes as inputs to the neural network. While the results generated by NNfoil-C during pretraining were satisfactory, achieving continuous solutions across a broad state-space.
Comparison with classical FVM results showed good agreement, highlighting the model's potential for efficient aerodynamic modeling under various flow conditions 
\medskip

\noindent \cite{10.1115/GT2024-122682} used PINNs along with generative artificial intelligence models to speed up and improve the accuracy of the compressor blade design process. The study solved a 2D incompressible viscous laminar N-S equation  for distinct turning angle values using a Fourier Projection layer to process inputs. A novel loss balancing technique, Relative Loss Balancing with Random Lookback (ReLoBRaLo), was proposed to improve training stability. The results for lift and drag closely matched CFD outputs and the hyperparameter selection including adaptive activation functions, learning rate scheduling, and signed distance functions significantly influencing training performance. The study also found that reducing point cloud density led to a reduction in the prediction error, indicating that adaptive sampling techniques could further enhance PINNs accuracy.
\medskip

\noindent \cite{doi:10.2514/1.J063562} proposed a model to predict the flow fields within an incompressible compressor cascade using PINNs.  The proposed method employs an adaptive weighting strategy to address the imbalance in the gradient computation during training and an adaptive learning rate scheme to enhance convergence. PINNs were applied to approximate two-dimensional subsonic linear compressor cascade flows, solving both forward and inverse problems. Forward predictions, based on CFD-derived boundary conditions, closely matched RANS results, while inverse modeling demonstrated the ability of PINNs to reconstruct flow fields from sparse boundary data.
The approach outperformed DNN in scenarios with limited training data and showed reduced MSE when accounting for aleatory uncertainty. However, the study also noted that PINNs still rely on CFD-generated data during training, limiting their ability to fully replace traditional solvers in industrial practice.
\subsection{Multi-Physics Interactions}

\noindent PINNs are increasingly being explored to address complex multi-physics problems in gas turbines, such as coupling between fluid dynamics, structural mechanics, and heat transfer. \cite{coulaud2024investigationsphysicsinformedneuralnetworks} investigated the adaptability of PINNs for parametric surrogate modeling, multi-physics coupling, and turbulence modeling via data assimilation. The study applied PINNs to model the flow field in a heated square cavity with varying viscosity and thermal conductivity, introducing flow parameters as additional inputs to the neural network. While PINNs successfully reconstructed the temperature field in conjugate heat transfer simulations, minor discrepancies were observed at the coupling interface when compared to finite element solutions.
Additionally, PINNs were tested on inverse problems in turbulence modeling, where they inferred turbulent viscosity fields from observed data, demonstrating potential in data-driven turbulence modeling. 
\medskip

\noindent \cite{10.1115/GT2024-128885} discusses the applicability of PINNs to interpret the flow of turbine blades in a transonic cascade using compressible inviscid equations. 
The study examined flow behavior over three turbine blade profiles under various training scenarios: with experimental data, without experimental data, and in an inverse setting without pressure boundary conditions, and compared the results with CFD simulations. PINNs was able to predict the discrepancies behind the trailing edge exhibited by the wedge region. The study also highlighted the sensitivity of PINNs to training data placement and optimization methods. For inverse problems, PINNs accurately predicted inlet and outlet Mach numbers based on static pressure data. While the study demonstrated the potential of PINNs in transonic cascade flow reconstruction, its applicability was limited to low $Re$ and low turbulence intensity.
\medskip

\noindent \cite{warey2021investigationnumericaldiffusionaerodynamic} demonstrated that PINNs can effectively mitigate numerical diffusion, a common issue in conventional CFD solvers when flow direction is misaligned with mesh lines. The study evaluated PINNs by solving steady-state incompressible and inviscid air convection problems at different flow angles, and accurately captured flow characteristics without artificial mixing layers. These findings suggest that PINNs offer a promising alternative for enhancing aerodynamic simulations, particularly in addressing numerical diffusion challenges, which remain a limitation in conventional CFD approaches.
\subsection{Real-Time Applications and Digital Twins}

\noindent PINNs are increasingly being explored for real-time applications in gas turbines, including digital twins and predictive maintenance, to enable advanced monitoring and prognostics. \cite{liu2022towards} proposed a hybrid physics and data-driven model to simulate the dynamic and transient performance of heavy-duty gas turbines (HDGTs). The primary objective was to develop an accurate and efficient simulation tool to facilitate fault diagnosis and predictive maintenance across a range of operating conditions. A thermodynamic model of a single-shaft gas turbine simulated the dynamic behavior of the gas turbine under varying conditions, integrating subsystems such as the compressor, combustor, turbine, rotating shaft, and control unit. To capture the transient behavior during the start-up process, a type of RNN model called nonlinear autoregressive with exogenous input (NARX) model, was employed. Bayesian regularization improved generalization, and the model achieved high prediction accuracy with a correlation above $0.98$. While effective in simulating both steady and transient states, the framework's reliance on high-quality data and long training times for the NARX component presents practical challenges. Nevertheless, the proposed methodology has the potential to be extended to real-time anomaly detection and fault diagnosis by incorporating timely operational data. 
\medskip

\noindent \cite{10.1115/1.4063326} applied PINNs to model an elastohydrodynamic (EHD) seal designed for supercritical carbon dioxide applications in high-pressure gas turbine systems. The study coupled a simplified Reynolds equation to describe fluid flow and Lame's equation to calculate structural deformation. Compared to a conventional prediction correction (PC) algorithm, PINNs demonstrated superior performance, accurately solving flow problems across a wide range of working pressures with greater flexibility. The results revealed a linear decrease in the pressure field for lower working pressures and a nonlinear decrease for higher pressures from inlet to outlet. The model accurately captured pressure and clearance variations across operating ranges and demonstrated the potential of PINNs for leakage prediction and seal optimization. This approach highlights the applicability of PINNs in modeling coupled fluid–structure phenomena under extreme operating conditions relevant to gas turbines. 
\medskip

\noindent \cite{BADORA2023102232} proposed a novel PINNs model to predict fatigue crack propagation in high-pressure gas turbine nozzles, using a limited dataset. The study aimed to enhance the reliability and lifespan of critical gas turbine components by accurately forecasting crack growth. The model, based on Paris' law, integrated a RNN structure with components for stress intensity factor estimation, cyclic loading, and time dependence. Each component contributed to the calculation of factors influencing crack growth. The model was trained on a dataset of 13 observations, incorporating visual inspection data, operational data from the Monitoring and Diagnostic System (M\&D), material properties, and cooling effectiveness coefficients derived from Finite Element Analysis (FEA). Compared to traditional data-driven models such as AdaBoost, Random Forest Regression, and conventional Neural Networks, the PINN approach achieved a normalized error of 9$\%$, even with small training datasets. Additionally, the model exhibited strong extrapolation capabilities and adaptability through a  Weibull inspired scaling method. While computational cost remains a concern, the approach demonstrates the promise of PINNs for predictive maintenance and structural integrity assessment in gas turbines.
\medskip

\noindent \cite{MARIAPPAN2024109388} developed an inverse PINN framework to study thermoacoustic interactions in a bluff-body anchored flame combustor. The framework integrates experimental data with a low-order model, composed of an acoustic equation and a Van der Pol oscillator equation to describe vortex shedding. A hybrid neural network was developed to reconstruct acoustic pressure and velocity fields, estimate heat release rates, and infer low-order model parameters. To address limited data availability, the study incorporated enhanced loss functions, self-adaptive weighting, and temporal segmentation, allowing for segment-wise prediction and stitching of time-varying acoustic responses. While the model successfully reconstructed thermoacoustic flows, it exhibited limitations in handling segmented time-series data and identifying missing physical dynamics in low-order models, suggesting the need for further refinements to enhance robustness and accuracy.

\begin{table}[!ht]
\caption{Summarised table of PINNs used for solving aerodynamic problems}
\label{tab1}
\begin{adjustbox}{width=\columnwidth,center}
\begin{tabular}{lllll}
\hline
\multicolumn{1}{c}{Reference} &
  \multicolumn{1}{c}{Focus Area} &
  \multicolumn{1}{c}{Equation} &
  \multicolumn{1}{c}{Methodology} &
  \multicolumn{1}{c}{Inference} \\ \hline
\cite{MAO2020112789} &
  \begin{tabular}[c]{@{}l@{}}High-Speed \\ Compressible Flows\end{tabular} &
  Euler equations &
  \begin{tabular}[c]{@{}l@{}}PINNs trained for forward and inverse \\ problems in 1D/2D high-speed flows.\end{tabular} &
  \begin{tabular}[c]{@{}l@{}}Accurate for smooth flows; struggled with \\ shock discontinuities; required clustered \\ sampling for steep gradients.\end{tabular} \\ \hline
\cite{10.1063/5.0095270} &
  \begin{tabular}[c]{@{}l@{}}RANS equations for \\ Turbulent Flows\end{tabular} &
  RANS equation &
  \begin{tabular}[c]{@{}l@{}}PINNs used to solve RANS equations \\ without turbulence model assumptions; \\ sparse boundary data used for training.\end{tabular} &
  \begin{tabular}[c]{@{}l@{}}Accurate for low-to-moderate $Re$ cases; \\ struggled with high-$Re$ turbulent flows \\ requiring better loss balancing.\end{tabular} \\ \hline
\cite{CAI20251} &
  \begin{tabular}[c]{@{}l@{}}Pressure Estimation from \\ Velocity\end{tabular} &
  N-S equation &
  \begin{tabular}[c]{@{}l@{}}PINNs trained to infer pressure \\ distributions from velocity data.\end{tabular} &
  \begin{tabular}[c]{@{}l@{}}Accurate for steady-state cases; needed \\ better loss balancing for transient flows.\end{tabular} \\ \hline
  \cite{10.1007/978-3-030-59851-8_9} &
  \begin{tabular}[c]{@{}l@{}}Parameter Identification \\ of RANS Turbulence \\ Models\end{tabular} &
  \begin{tabular}[c]{@{}l@{}}$k-\epsilon$ turbulence \\ equation\end{tabular} &
  \begin{tabular}[c]{@{}l@{}}Physics-embedded neural networks used \\ to infer unknown turbulence model \\ parameters.\end{tabular} &
  \begin{tabular}[c]{@{}l@{}}Helped improve RANS predictions; \\ limited to canonical flows.\end{tabular} \\ \hline
  \cite{LI2022114524} &
  \begin{tabular}[c]{@{}l@{}}Reynolds Equation for \\ Gas Bearings\end{tabular} &
  Reynolds Equation &
  \begin{tabular}[c]{@{}l@{}}PINNs used to solve lubrication \\ flow problems in gas bearings.\end{tabular} &
  \begin{tabular}[c]{@{}l@{}}Good accuracy for thin-film flows; \\ needed refinement for transient cases.\end{tabular} \\ \hline
\cite{fluids8020043} &
  Turbulence Modeling &
  RANS equation &
  \begin{tabular}[c]{@{}l@{}}Comparison of different turbulence \\ models for backward-facing step flow \\ using PINNs.\end{tabular} &
  \begin{tabular}[c]{@{}l@{}}Effective in low-$Re$ flows; high \\ $Re$ showed increased error.\end{tabular} \\ \hline
\cite{HANRAHAN2023109232} &
  \begin{tabular}[c]{@{}l@{}}Turbulent Flows with\\ Sparse Data\end{tabular} &
  \begin{tabular}[c]{@{}l@{}} RANS equation\end{tabular} &
  \begin{tabular}[c]{@{}l@{}}PINNs trained with minimal \\ experimental data to reconstruct turbulent \\ velocity and pressure fields.\end{tabular} &
  \begin{tabular}[c]{@{}l@{}}Showed promising turbulence \\ reconstruction; sensitivity to training \\ data placement; high $Re$ flows needed \\ additional constraints.\end{tabular} \\ \hline
\cite{LIU2024113275} &
  \begin{tabular}[c]{@{}l@{}}Reconstruction of \\ Turbulent Flames from \\ Sparse Data\end{tabular} &
  \begin{tabular}[c]{@{}l@{}}Combustion and \\ Heat Transfer Equations\end{tabular} &
  \begin{tabular}[c]{@{}l@{}}PINNs used to infer flame \\ structures from sparse experimental \\ data.\end{tabular} &
  \begin{tabular}[c]{@{}l@{}}Accurate for steady flames; \\ transient effects required additional \\ constraints.\end{tabular} \\ \hline
  \cite{10.1115/GT2023-101238} &
  \begin{tabular}[c]{@{}l@{}}Transitional and Turbulent \\ Flow around Turbine Blades\end{tabular} &
  \begin{tabular}[c]{@{}l@{}}N-S equation\end{tabular} &
  \begin{tabular}[c]{@{}l@{}}PINNs applied to periodic hills and \\ T106C turbine blade cases; trained \\ with sparse data for turbulence closure.\end{tabular} &
  \begin{tabular}[c]{@{}l@{}}Good agreement with CFD; errors \\ at leading/trailing edges due to sparse \\ data; high computational cost for \\ compressible flow.\end{tabular} \\ \hline
\cite{10.1063/5.0188665} &
  \begin{tabular}[c]{@{}l@{}}Subsonic Flow \\ Around Airfoils\end{tabular} &
  \begin{tabular}[c]{@{}l@{}}Potential Flow \\ Equations\end{tabular} &
  \begin{tabular}[c]{@{}l@{}}PINNs with mesh transformation \\ to solve subsonic flow over airfoils.\end{tabular} &
  \begin{tabular}[c]{@{}l@{}}Performed well in low-speed regimes; \\ required adjustments for high-lift \\ configurations.\end{tabular} \\ \hline
\cite{CAO2024113285} &
  \begin{tabular}[c]{@{}l@{}}Parametric Engineering \\ Problems for \\ Inviscid Flows\end{tabular} &
  Euler Equations &
  \begin{tabular}[c]{@{}l@{}}PINNs trained for parametric inviscid \\ airfoil flows using high-dimensional \\ embedding.\end{tabular} &
  \begin{tabular}[c]{@{}l@{}}Scalable for design studies; lacked \\ viscous effects modeling.\end{tabular} \\ \hline
\cite{10.1115/GT2024-122682} &
  \begin{tabular}[c]{@{}l@{}}Airfoil Design \\ Optimization\end{tabular} &
  \begin{tabular}[c]{@{}l@{}}N-S equation\end{tabular} &
  \begin{tabular}[c]{@{}l@{}}PINNs applied to airfoil shape \\ optimization.\end{tabular} &
  \begin{tabular}[c]{@{}l@{}}Demonstrated aerodynamic shape \\ improvements; lacked experimental \\ validation.\end{tabular} \\ \hline
  \cite{doi:10.2514/1.J063562} &
  \begin{tabular}[c]{@{}l@{}}Compressor Cascade Flow \\ with Adaptive Learning\end{tabular} &
  RANS Equations &
  \begin{tabular}[c]{@{}l@{}}Adaptive PINNs strategy applied \\ to predict compressor cascade \\ aerodynamics.\end{tabular} &
  \begin{tabular}[c]{@{}l@{}}Adaptive learning improved convergence; \\ struggled with secondary flow effects.\end{tabular} \\ \hline
\cite{coulaud2024investigationsphysicsinformedneuralnetworks} &
  \begin{tabular}[c]{@{}l@{}}Multi-Physics Coupling, \\ Surrogate Modeling, \\ Turbulence Modeling\end{tabular} &
  \begin{tabular}[c]{@{}l@{}}N-S, Conjugate Heat \\ Transfer Equations\end{tabular} &
  \begin{tabular}[c]{@{}l@{}}PINNs used for parametric surrogate \\ modeling, inverse turbulence \\ modeling, and multi-physics coupling.\end{tabular} &
  \begin{tabular}[c]{@{}l@{}}Accurately modeled conjugate heat \\ transfer and turbulence fields; some \\ discrepancies at coupling interfaces \\ and optimizer dependency.\end{tabular} \\ \hline
  \cite{10.1115/GT2024-128885} &
  \begin{tabular}[c]{@{}l@{}}Flow Reconstruction in \\ Transonic Turbine \\ Cascades\end{tabular} &
  N-S Equations &
  \begin{tabular}[c]{@{}l@{}}PINNs used to infer missing \\ flow field data in a transonic \\ turbine cascade.\end{tabular} &
  \begin{tabular}[c]{@{}l@{}}Good reconstruction accuracy; \\ required high-quality sensor data for \\ training.\end{tabular} \\ \hline

\cite{warey2021investigationnumericaldiffusionaerodynamic} &
  \begin{tabular}[c]{@{}l@{}}Numerical Diffusion in \\ Aerodynamic Flow \\ Simulations\end{tabular} &
  N-S Equation &
  \begin{tabular}[c]{@{}l@{}}PINNs studied for minimizing \\ numerical diffusion effects in \\ flow solvers.\end{tabular} &
  \begin{tabular}[c]{@{}l@{}}Showed promise in stabilizing flow \\ solutions; computationally expensive.\end{tabular} \\ \hline

\cite{liu2022towards} &
  \begin{tabular}[c]{@{}l@{}}Predictive Maintenance \\ of Heavy-Duty Gas \\ Turbines\end{tabular} &
  \begin{tabular}[c]{@{}l@{}}Performance Simulation\\ Models\end{tabular} &
  \begin{tabular}[c]{@{}l@{}}Hybrid AI-PINNs framework \\ applied for gas turbine condition \\ monitoring.\end{tabular} &
  \begin{tabular}[c]{@{}l@{}}Effective for fault prediction; \\ required domain-specific tuning.\end{tabular} \\ \hline
\cite{10.1115/1.4063326} &
  \begin{tabular}[c]{@{}l@{}}Elastohydrodynamic Seals \\ for Supercritical CO2 \\ Turbomachinery\end{tabular} &
  \begin{tabular}[c]{@{}l@{}}Reynolds equation, \\ Lame's formula\end{tabular} &
  \begin{tabular}[c]{@{}l@{}}Deep learning-based PINNs applied \\ to novel turbomachinery seals.\end{tabular} &
  \begin{tabular}[c]{@{}l@{}}Showed feasibility for design; \\ lacked experimental validation.\end{tabular} \\ \hline
\cite{BADORA2023102232} &
  \begin{tabular}[c]{@{}l@{}}Predicting Gas Turbine \\ Nozzle Cracks\end{tabular} &
  \begin{tabular}[c]{@{}l@{}}Fracture Mechanics \\ Equations\end{tabular} &
  \begin{tabular}[c]{@{}l@{}}Small dataset PINNs trained to estimate \\ crack lengths in turbine nozzles.\end{tabular} &
  \begin{tabular}[c]{@{}l@{}}Performed well with limited data; \\ needed better uncertainty quantification.\end{tabular} \\ \hline
\cite{MARIAPPAN2024109388} &
  \begin{tabular}[c]{@{}l@{}}Thermoacoustic \\ Interactions in Combustors\end{tabular} &
  \begin{tabular}[c]{@{}l@{}}N-S, Acoustic \\ Wave Equations\end{tabular} &
  \begin{tabular}[c]{@{}l@{}}PINNs trained to capture \\ thermoacoustic instabilities in gas \\ turbine combustors.\end{tabular} &
  \begin{tabular}[c]{@{}l@{}}Identified instability growth rates; \\ struggled with high-frequency oscillations.\end{tabular} \\ \hline
\end{tabular}
\end{adjustbox}
\end{table}
\section{PINNs for aeromechanics}\label{sec5}
\noindent The aeromechanical characteristics of gas turbine components involve the complex interplay between structural dynamics and aerodynamic forces, often referred to as Collar's aeroelastic triangle. This interaction comes with key phenomena such as flutter, and forced response ~(\cite{app131911079,9670772,aerospace8090242}). Accurate modeling of these phenomena is critical for preventing fatigue failure and ensuring turbine reliability. In response to these challenges, DL techniques have been increasingly applied to address these challenges~(\cite{WeiZhengYanLiChiJiang+2025+99+114,10.1063/5.0183290,10314851}), offering data-driven, physics-informed, and computationally efficient solutions. Among these, PINNs have gained traction as a promising framework for solving specific aeromechanical problems. While only a limited number of studies have been conducted in this area, we have highlighted a selection that closely aligns with our survey scope, as summarized in Table \ref{tab2}, focusing on their methodologies and key features.

\medskip
\noindent One notable application is the use of PINNs to model the complex dynamics of mistuned bladed disks in turbomachinery was presented by \cite{10.1115/GT2024-125397}. A physics-informed surrogate model, termed the Designed Structural (DS) neural network, was developed to capture the effects of random mistuning patterns in blade properties. By embedding cyclic symmetry shift features directly into the network architecture, the model significantly reduced the need for large training datasets. It achieved high predictive accuracy for blade displacement under mistuning, with an $R^2$ value of  $0.99$, and an amplitude magnification error of $0.68\%$. The proposed surrogate model successfully predicts the vibrational response, offering a computationally efficient alternative to conventional FEA. This approach holds promise for real-time monitoring and control applications in turbomachinery, where understanding the dynamic behavior of components is crucial for ensuring operational safety and longevity.\medskip

\noindent Another attempt by \cite{Rauseo2024} introduced a Physics-Guided Machine Learning (PGML) framework to predict flutter stability in compressor cascades, combining machine learning with reduced order models (ROMs). This approach incorporates prior knowledge of flutter phenomena by formulating relevant steady-state input features and injecting results from low-fidelity analytical models into the learning process, eliminating the need for computationally intensive unsteady simulations. This efficiency allows for rapid assessment of flutter stability across different blade geometries and mechanical properties, such as mode shapes and frequencies, with minimal additional computational cost once the mean flow is known. In particular, the model demonstrated strong generalization, achieving accurate predictions even when trained on a single-compressor geometry, highlighting its potential for efficient and scalable flutter analysis in turbomachinery.\medskip

\noindent \cite{2021CompM..67..619Z} proposed a PINN framework, one among the earliest studies applying physics-informed deep learning to three-dimensional metal Additive Manufacturing (AM). This approach ensures that the model adheres to established physical principles, enhancing prediction accuracy even with limited data. The method was based on the heaviside function to impose Dirichlet boundary conditions within the neural network. This "hard" enforcement not only guarantees precise adherence to boundary conditions but also accelerates the training process compared to traditional "soft" constraint methods. The PINN framework was validated using the $2018$ NIST AM Benchmark Test Series~(\cite{levine2020outcomes}) and demonstrated strong agreement with both experimental and high-fidelity simulation data. It is also shown that the model outperforms traditional ML models that often require large labeled datasets, which are expensive to obtain in AM due to the high cost of experiments and computationally intensive simulations.\medskip

\noindent Zhang et al.~\cite{ZHANG2021108130} presents an innovative approach to predict the creep-fatigue life of 316 stainless steel. The authors enhanced the input features of the neural network by incorporating domain-specific physical knowledge. This process involves deriving features that capture the underlying mechanics of creep and fatigue phenomena, thereby improving the model's ability to generalize from limited data. A novel loss function is formulated to ensure that the model's predictions adhere to established physical principles, enhancing both accuracy and reliability. Compared to traditional ML and standard deep neural networks, the PINN approach demonstrated superior generalization and  highlights the potential of PINNs in enhancing material life prediction accuracy while ensuring adherence to physical laws.\medskip

\noindent \cite{Xie_Zhao_Zhao_Fu_Shelton_Semlitsch_2024} proposed a PINN-based model for capturing thermoacoustic instabilities in combustion systems, focusing on bifurcation behavior and amplitude death characteristics. By extending the classical Van der Pol oscillator and integrating it with a nonlinear Rijke-type combustor model, the framework accurately predicted frequency, limit cycle amplitude, and transition times to large-amplitude oscillations. The PINN approach successfully identified bifurcation points and conditions leading to amplitude death which are crucial for understanding and mitigating transitions to instability in thermoacoustic systems. \medskip

\noindent \cite{vermaphysics} explored the application of PINNs in structural mechanics by solving the Euler-Bernoulli equation for a cantilever beam subjected to uniformly distributed loads. 
The study evaluated the impact of various hyperparameters, including activation functions, optimizers, and training iterations, on convergence and accuracy. With Adam optimizer, tanh activation, and $2000$ epochs, the PINNs model closely matched the analytical solution.
Furthermore, the training process took comparatively less time, demonstrating the efficiency of PINNs in terms of computational time. These findings highlight the potential of data-driven physics-informed approaches for applications in digital twins for aerospace and mechanical engineering systems, demonstrating PINNs efficiency and accuracy in structural mechanics simulations.\medskip

\noindent \cite{modelling5040080} applied PINNs to model the deflection of a one-dimensional prismatic cantilever beam, representative of helicopter blade structures, under triangular loading. The governing fourth-order differential equation was solved using a neural network with $5$ hidden layers and $50$ neurons per layer, trained on $51$ collocation points. Results from the PINNs model demonstrated close agreement with analytical solutions, outperforming a standard Artificial Neural Network (ANN) model of identical architecture. Furthermore, PINNs accurately predicted the deflection, slope, bending moment, and shear force of the beam, aligning precisely with the exact results. The MSE values for these predicted quantities were significantly lower compared to the ANN results.
Moreover, the computational time and cost for PINNs were reduced compared to classical numerical methods.
These findings highlight the promise of PINNs for efficient and accurate structural analysis in aerospace applications.\medskip

\noindent \cite{10255379} proposed a PINN-based methodology for solving non-dimensional single and double beam systems, under both forward and inverse settings. To address convergence issues caused by large coefficients in the dimensional Euler-Bernoulli formulation, the study adopted a non-dimensional approach for enhanced numerical stability. For the Timoshenko beam forward problem, PINNs successfully predicted transverse displacement and cross-sectional rotation with relative errors below $10^{-3}\%$, respectively. In the inverse problem, PINNs were used to estimate internal material properties and unknown force functions, achieving less than $3\%$ error, demonstrating potential for structural health monitoring and early failure detection. For Euler-Bernoulli double beam systems connected by a Winkler foundation, PINNs accurately predicted velocity, bending moment, and acceleration. Comparisons with Finite Difference Methods (FDM), Physics-Guided Neural Networks (PGNNs), and gradient-enhanced PINNs (gPINNs) demonstrated that PINNs outperformed these methods, suggesting that higher model complexity does not necessarily reduce prediction accuracy. These results underscore their effectiveness in structural mechanics applications, particularly in scenarios with limited or noisy data.

\medskip

\begin{table}[!ht]
\caption{Summarised table of PINNs used for solving aeromechanic problems}
\label{tab2}
\begin{adjustbox}{width=\columnwidth,center}
\begin{tabular}{lllll}
\hline
\multicolumn{1}{c}{Reference} &
  \multicolumn{1}{c}{Focus Area} &
  \multicolumn{1}{c}{Equation} &
  \multicolumn{1}{c}{Methodology} &
  \multicolumn{1}{c}{Inference} \\ \hline
\cite{10.1115/GT2024-125397} &
  \begin{tabular}[c]{@{}l@{}}Mistuned Bladed \\ Disks Dynamics\end{tabular} &
  \begin{tabular}[c]{@{}l@{}}Structural Dynamics \\ Equations\end{tabular} &
  \begin{tabular}[c]{@{}l@{}}PINN surrogate model for mistuned \\ blade vibrations in gas turbines.\end{tabular} &
  \begin{tabular}[c]{@{}l@{}}Accurate in predicting mistuning effects; \\ struggled with nonlinear coupling at \\ higher harmonics.\end{tabular} \\ \hline
\cite{Rauseo2024} &
  \begin{tabular}[c]{@{}l@{}}Compressor Stall and \\ Flutter\end{tabular} &
  \begin{tabular}[c]{@{}l@{}}Aeroelasticity \& Fluid-Structure \\ Interaction Equations\end{tabular} &
  \begin{tabular}[c]{@{}l@{}}Physics-guided machine learning \\ to model stall-induced flutter in \\ compressors.\end{tabular} &
  \begin{tabular}[c]{@{}l@{}}Captured stall onset well; required \\ high-quality data for training to improve \\ robustness.\end{tabular} \\ \hline
\cite{2021CompM..67..619Z} &
  Metal Additive Manufacturing &
  \begin{tabular}[c]{@{}l@{}}Heat Transfer \& Fluid \\ Dynamics Equations\end{tabular} &
  \begin{tabular}[c]{@{}l@{}}PINNs used to predict temperature and \\ melt pool dynamics in additive \\ manufacturing.\end{tabular} &
  \begin{tabular}[c]{@{}l@{}}Demonstrated strong agreement with\\ both experimental and high-fidelity\\ simulation results; outperformed ML \\methods under scarce data.\end{tabular} \\ \hline
\cite{ZHANG2021108130} &
  Creep-Fatigue Life Prediction &
  \begin{tabular}[c]{@{}l@{}}Fracture Mechanics \& \\ Material Degradation Models\end{tabular} &
  \begin{tabular}[c]{@{}l@{}}PINNs used to predict long-term \\ creep-fatigue behavior in \\ high-temperature components.\end{tabular} &
  \begin{tabular}[c]{@{}l@{}}Accurate for smooth creep transitions; \\ needed refined training for sharp fatigue \\ failures.\end{tabular} \\ \hline
\cite{Xie_Zhao_Zhao_Fu_Shelton_Semlitsch_2024} &
  Thermoacoustic Instabilities &
  Van der Pol Oscillator Model &
  \begin{tabular}[c]{@{}l@{}}PINNs used to analyze bifurcation and \\ amplitude death in thermoacoustic \\ instability systems.\end{tabular} &
  \begin{tabular}[c]{@{}l@{}}Good agreement with experimental trends; \\ struggled with highly nonlinear limit-cycle \\ behaviors.\end{tabular} \\ \hline
\cite{vermaphysics} &
  \begin{tabular}[c]{@{}l@{}}Computational Structural \\ Mechanics\end{tabular} &
  Solid Mechanics Equations &
  \begin{tabular}[c]{@{}l@{}}PINNs applied for solving computational \\ structural mechanics problems.\end{tabular} &
  \begin{tabular}[c]{@{}l@{}}Accurate in low-strain conditions; \\ required enhanced constraints for complex \\ material behaviors.\end{tabular} \\ \hline
\cite{modelling5040080} &
  1D Solid Mechanics &
  1D Elasticity Equations &
  \begin{tabular}[c]{@{}l@{}}PINNs applied to solve forward and inverse \\ problems in 1D structural mechanics.\end{tabular} &
  \begin{tabular}[c]{@{}l@{}}Efficient for simple cases; struggled with \\ high-deformation scenarios.\end{tabular} \\ \hline
\cite{10255379} &
  Beam Systems &
  \begin{tabular}[c]{@{}l@{}}Beam Deflection and Stress \\ Equations\end{tabular} &
  \begin{tabular}[c]{@{}l@{}}PINNs used for solving forward and inverse \\ problems in complex beam structures.\end{tabular} &
  \begin{tabular}[c]{@{}l@{}}Accurate for static loading; needed \\ improvements for dynamic response \\ predictions.\end{tabular} \\ \hline
\end{tabular}
\end{adjustbox}
\end{table}

\section{Recent Advances in PINNs for Aerodynamics and Aeromechanics}\label{sec6}

\noindent PINNs have seen rapid development in recent years, finding novel applications in the aerodynamic and aeromechanical domains of IGTs. This section summarizes recent advancements in the PINNs framework to real-world engineering problems that offers a strong foundation for further exploration in gas turbine systems.\medskip

\noindent Several improvements in the PINNs framework over the past years have enhanced their applicability to complex, multi-physics problems like those in gas turbines. One major focus has been convergence and training efficiency for PINNs, which historically can suffer from slow or unstable training on stiff PDE~(\cite{sharma2023stiff}) or long-time simulations. Techniques such as adaptive loss weighting, learning rate scheduling, and activation function tuning have shown to significantly speed up convergence.
One such approach is extending PINNs capabilities with multi-task learning and transfer learning~(\cite{XU2023115852}) to tackle inverse structural mechanics problems. This improves efficiency and accuracy when applying PINNs to inverse problems in structural mechanics, where unknown external forces must be identified. The PINNs was initially trained on a simplified scenario, then fine-tuned for a target scenario, which greatly accelerates convergence and improves accuracy. The study demonstrated that the PINN can successfully characterize the unknown loads of a structural system for sparse and noisy observational data. The physics-based regularization in PINNs acted as a form of built-in prior, making the model less sensitive to noise and more robust than classical inverse methods. Also, the non-dimensionalization and utilization of homoscedastic uncertainty to weigh different learning tasks significantly improved the performance of PINNs. The authors put forth a few challenges regarding handling very localized loads that cause sharp stress gradients and suggest that advanced network architectures or neural operators could further improve PINN performance in those cases. Nevertheless, the presented results exhibit how PINNs augmented with smart training strategies can bridge the gap between data-driven models and physics-based simulation, offering fast, reliable surrogates for engineering analysis.\medskip

\noindent To enhance the performance of PINNs for complex multi-physics problems and to scale PINNs to realistic $3$D turbine geometries and high-fidelity simulations, two novel frameworks were introduced: conservative PINNs (cPINNs)~(\cite{JAGTAP2020113028}) and extended PINNs (XPINNs)~(\cite{osti_2282003}). cPINNs address the limitations of classical PINNs by employing non-linear conservative laws within discrete, subdomain-fragmented domains. By enforcing rigid continuity between subdomains and averaging solutions at subdomain interfaces, cPINNs ensure the conservative properties of the solution. This domain decomposition approach inherently aligns with parallel computation paradigms, where each subdomain can be treated as an independent computational unit.
Building upon cPINNs, XPINNs further extend the capabilities of PINNs by offering a generalized architecture that combines the strengths of both classical and conservative approaches. XPINNs utilize spatio-temporal domain decomposition techniques, enabling the solution of nonlinear PDE on arbitrary complex geometries. Unlike standard PINNs, XPINNs exhibit improved representation and parallelization capabilities, facilitated by the use of distinct, customizable neural networks for each subdomain. A key innovation of XPINNs is its adaptability to any PDE, along with its flexible domain decomposition, which enables both spatial and temporal parallelization, leading to streamlined training. This flexibility allows for the tailoring of neural network architectures and optimization parameters to specific domain characteristics. Both cPINNs and XPINNs leverage domain decomposition to enhance parallelization, making them highly generalizable in terms of domain applicability and decomposition, thereby offering more efficient and robust solutions for complex scientific computing problems.\medskip

\noindent Another advancement introduces a PINNs model specifically designed to solve the inverse heat conduction problem~(\cite{10.1115/1.4067125}) in the rotating cavity of a high-pressure compressor system. Inverse heat transfer problems in gas turbines are notoriously challenging due to the ill-posed nature of inferring heat flux from limited temperature measurements. This framework learns to map experimentally measured radial temperature distributions to the unknown surface heat fluxes on the rotating cavity walls. By combining experimental data and fundamental heat conduction physics in the training process, the PINNs framework successfully learns the relationship between internal disk temperature profiles and surface heat fluxes. Key outcomes include the ability to rapidly predict heat fluxes that, when input into a standard solver, reproduce the measured temperatures with good fidelity. This confirms the PINNs effectiveness in solving the inverse problem and validating the results against known physics. The framework is robust against noise, a critical advantage in practical scenarios and represents a noteworthy advancement over classical inverse methods, which often struggle with experimental uncertainties. The proposed PINNs based method advances the state-of-the-art by offering a fast, accurate, and more noise-resilient tool for interpreting thermal measurements in gas turbine cavities.\medskip

\noindent One of the critical challenges faced by PINNs is the model misspecification problem, which arises when the underlying physics used to define the loss function is inaccurate or incomplete. This can occur because the governing equations are approximations, omit key phenomena, or rely on assumptions that do not hold under all conditions. As a result, even a well-trained PINN may produce biased or unreliable predictions due to learning from an incorrect or oversimplified physical model. To address this problem, a hybrid framework was proposed by~(\cite{ZOU2024112918}), which combines standard PINNs with additional DNNs to correct discrepancies between the assumed and actual physics. The method also integrates Bayesian PINNs (B-PINNs)~(\cite{YANG2021109913}) and ensemble PINNs to quantify uncertainties in the discovered governing equations. It also employs symbolic regression to extract explicit mathematical expressions for the missing physics, enhancing interpretability. By leveraging uncertainty quantification and data-driven corrections, this framework enables PINNs to generalize better across scientific domains where physics models are incomplete.\medskip

\noindent In the paper \cite{naujoks2024pinnfluenceinfluencefunctionsphysicsinformed}, the authors employed a novel approach to PINNs by applying influence functions (IFs) to analyze and evaluate the role of collocation points on the model's behavior and prediction accuracy. This approach was demonstrated through the application of the N-S equations to a laminar, incompressible, and time-dependent fluid flow.
The experiments were conducted using three distinct models: a well-trained model that yielded excellent results compared to the reference solution, an under-trained model, and a model with incorrect physics, neither of which produced accurate results. For empirical analysis, two heuristically motivated IF indicators were used: Directional Indicators (DI) to capture the physical aspects of the PDE and their influence on PINNs, and Region Indicators (RI) to identify regions that strongly influence other training points.
The values of DI reveal the model's ability to learn the underlying physics in the direction of flow. The results obtained demonstrate that IFs effectively reveal the impact of training points near the region around the cylinder and on the boundary.  The DI and RI metrics also showed excellent results for the well-trained model, demonstrating its close alignment with the indicators' assumptions.
Hence, this study highlights the potential benefits of IFs as a valuable post-hoc tool for debugging and validating PINNs, thereby enhancing their interpretability and reliability.\medskip

\noindent The paper by \cite{qiu2024pifusionphysicsinformeddiffusionmodel} introduces a novel methodology to predict the temporal evolution of velocity and pressure fields. This methodology utilizes a physics-informed deep learning method called Pi-fusion, which combines the advantages of physical laws and diffusion models capable of learning data distributions.
The model was demonstrated by learning the fluid dynamics of incompressible Newtonian flows. Initially, the model comprises a diffusion process where Gaussian noise is added to the fluid dynamics simulation. This is followed by a reverse process that reconstructs the velocity and pressure fields based on a physics-informed score function.
To enhance the interpretability of the model, a reciprocal learning-based strategy was employed to explore the quasiperiodic behavior of the fluid flow. Additionally, physics-informed guidance sampling was also employed to further encourage model interpretability.
This approach overcomes a limitation of traditional physics-informed deep learning models: their inability to generalize across arbitrary time instants. The approach has been validated on both synthetic and real-world problems and compared with existing state-of-the-art PINNs. Experimental results demonstrate that the proposed model outperforms others for both synthetic and real-world datasets. \medskip

\noindent PINNs are used to solve incompressible flow problems involving time-dependent moving boundaries~(\cite{10.1063/5.0186809}), overcoming the limitation of traditional PINNs, which are primarily designed for stationary boundary problems. An extended PINNs formulation enforces no-slip boundary conditions at moving interfaces through modified loss functions while refining the training point distribution near dynamic boundaries to enhance accuracy. The study demonstrates the effectiveness of this framework through multiple numerical experiments, including oscillating and orbiting cylinders, bio-inspired flapping wings, and vortex shedding in a uniform flow, validating its ability to reconstruct flow fields even with sparse sensor data. The results show that PINNs successfully capture complex hydrodynamic interactions and provide accurate predictions comparable to high-fidelity CFD simulations, making them a promising alternative for fluid-structure interaction problems, aerodynamics, and bio-inspired propulsion systems. Key advantages of the method include its ability to eliminate remeshing requirements, its scalability across different unsteady flow scenarios, and its robustness in reconstructing flow fields with limited data. However, challenges remain in terms of computational cost, uncertainty quantification, and handling more complex coupled multiphysics problems, which can be considered for future research.\medskip

\noindent Each of these studies highlights both the potential and current challenges of PINNs in industrial gas turbine applications. Improved training techniques like adaptive losses, hybrid models, and transfer learning, have enhanced their stability and speed, making them increasingly viable for real-world use. While traditional CFD and FEM remain the workhorses for detailed turbine analysis, PINNs are emerging as powerful complementary tool particularly for inverse problems and scenarios with limited data. Ongoing research and case studies continue to refine PINNs capabilities, bringing us closer to physics-informed AI that can reliably assist in the design and analysis of next-generation gas turbines.

\section{Challenges and Future directions}\label{sec7}

\noindent Despite the significant potential of PINNs for modeling aerodynamic and aeromechanical characteristics in IGTs, extensive research has revealed several challenges associated with their training and implementation. These challenges, while actively being addressed, can impact the accuracy, efficiency, and reliability of PINNs in IGTs applications. Some common issues include convergence difficulties, high computational costs, spectral bias, and limitations in handling data stiffness. Additionally, certain mathematical aspects of PINNs are still in their early stages, lacking comprehensive theoretical foundations and rigorous analysis of convergence and error estimation. This section delves into these major shortcomings and explores potential future research directions to enhance the applicability of PINNs for addressing complex problems in IGTs.
\begin{enumerate}
    \item \textbf{Sampling in High-Dimensional Spaces}: PINNs often require a very large number of training (collocation) points to capture complex flow solutions in high-dimensional domains. This requirement grows rapidly with problem dimension and solution complexity, leading to a “curse of dimensionality”~(\cite{NEURIPS2023_4af827e7}). This poses difficulties in selecting an appropriate distribution of collocation points, as uniform sampling may not effectively resolve localized flow features or turbulent eddies. Studies have shown that as problem dimensionality increases, conventional PINNs demand substantially more collocation points, leading to higher computational costs and memory usage. Furthermore, in the context of IGTs problems, PINNs may not yield accurate solutions without labeled data. For example, for high-Reynolds-number flows, fully connected feedforward PINNs have failed to produce acceptable predictions when trained without labeled data within the domain~(\cite{fluids8020043}). 
    \item \textbf{High Computational Cost and Inefficiency}: PINNs remain computationally expensive and often trained slowly, particularly when compared to traditional CFD or FEM solvers for forward problems. Each training iteration involves evaluating the PDE residuals at numerous collocation points via automatic differentiation, which can be significantly more time consuming than a single time step of a CFD solver. In one benchmark, a PINNs computation took 32 hours, while the CFD simulation took less than 20 seconds~(\cite{chuang2022experiencereportphysicsinformedneural}). This highlights that basic PINNs can be orders of magnitude slower than optimized approaches or classical methods. The computational cost scales poorly with the number of PDE constraints and points, for instance, increasing collocation points or enforcing additional physics significantly slows convergence. PINNs training is essentially a high-dimensional, non-convex optimization, and it may require many iterations to converge. Memory overhead is also substantial, as storing gradients for millions of points can exhaust GPU memory. This inefficiency currently limits their direct use as replacements for conventional solvers in industry.
    \item \textbf{Convergence and Accuracy in Complex Flows}: Ensuring the convergence of PINNs to accurate solutions, especially for turbulent or high-speed flows relevant to gas turbines, remains a challenge. Neural networks exhibit a spectral bias~(\cite{10630853}), learning low-frequency solution components faster than high-frequency details. Consequently, PINNs may struggle to represent fine turbulent eddies or sharp gradients, often learning these features very slowly or not at all. This can lead to significant errors in complex flow regions. Additionally, solving stiff or highly nonlinear PDE can cause numerical instabilities, requiring careful regularization and training strategies. Unlike traditional solvers with established stability criteria, PINNs may converge slowly or become trapped in local minima, especially without well-chosen network architectures and loss weighting. Recent studies suggest techniques like adaptive loss balancing, curriculum training, and improved activation functions to enhance stability and convergence speed.
    \item \textbf{Multi-Physics Coupling Difficulties}: IGTs problems inherently involve multiple physical phenomena such as fluid dynamics, structural mechanics, heat transfer, and often combustion chemistry, acting simultaneously. Incorporating all these physics into a single PINNs framework is challenging. Each additional physical field introduces extra PDE or constraints in the loss function, increasing the complexity of the optimization problem. Balancing these multiple objectives is difficult, as the network might prioritize one physics over another, leading to violations in the less weighted equations. In practice, PINNs can struggle to enforce all coupled physics constraints simultaneously with sufficient accuracy. Current experience shows that directly applying PINNs to strongly coupled multi-physics problems is computationally prohibitive and unstable, often requiring decomposition into sub-problems or some form of prior knowledge injection. This is a key obstacle to applying PINNs directly to full IGTs simulations, where aerodynamics and aeromechanics are deeply intertwined.
    \item \textbf{Scalability to Large-Scale Industrial Problems}: Solving PDE within the complex, 3D geometries of real-world gas turbines, such as combustors or turbine stages, pushes PINNs to their computational limits. Industrial problems often involve millions of spatial degrees of freedom and intricate boundaries, requiring traditional CFD meshes with millions of cells. Similarly, PINNs would need to handle vast domains or high dimensional input spaces to capture all relevant flow features. Training such large scale networks can become prohibitively expensive due to the poor scaling of memory and computational requirements. Without specialized techniques, training on large 3D domains with fine detail could take an impractical amount of time.Another challenge is that PINNs inherently solve for all points simultaneously, unlike traditional solvers, which march solutions locally, facilitating easier parallelization across multiple processors. While PINNs are mesh-free, offering advantages for complex geometries, this lack of spatial localization can hinder scalability. Although data-parallel training on multiple GPUs or distributed clusters is possible, communication and synchronization overheads can limit performance for extremely large models. Currently, conventional PINNs do not scale efficiently to the size of real engine components, often being limited to simpler or smaller domains. While efforts by industry, such as NVIDIA Modulus~(\cite{miao2024implementing}), have demonstrated PINNs applications on HPC clusters for large cases, these often require significant engineering and still face challenges with training times. Therefore, handling the full complexity of an IGTs with PINNs alone remains out of reach, and scalability continues to be a pressing challenge.
\end{enumerate}
\noindent While PINNs present significant potential for IGTs applications, several challenges remain. To address these, a promising strategy is combining PINNs with traditional modeling and multi-fidelity data. Hybrid approaches, such as PINNs-CFD integrations, leverage conventional solvers for partial information, allowing PINNs to fill gaps and enforce additional physics, effectively acting as physics-based interpolation tools constrained by both data and governing equations. Multi-fidelity learning~(\cite{PENWARDEN2022110844}), which trains PINNs on a blend of low-fidelity and sparse high-fidelity data, drastically reduces computational costs while maintaining accuracy.
To overcome training inefficiencies inherent in basic PINNs, researchers are developing advanced optimization and sampling techniques. Adaptive training, including adaptive sampling~(\cite{WU2023115671}) and loss weighting~(\cite{2023NonDy.11115233H,YU2022114823}), prevents the network from prioritizing easier aspects of the problem over harder constraints. Algorithms like Neural Tangent Kernels~(\cite{WANG2022110768}) enhance training stability, while second-order optimizers and advanced schedulers improve loss landscape navigation. Implementing generalized space-time domain decomposition further accelerates training and enhances accuracy by enabling the use of smaller networks for local behavior and limiting global communication to interfaces.
Ultimately, a key goal for PINNs development is to enable digital twins~(\cite{YANG2024117075}) of complex systems like gas turbines. PINNs, with their ability to blend data and physics, are ideally suited for real-time inference on streaming sensor data. Future research will focus on deploying pre-trained PINNs for real-time fine-tuning with incoming measurements, paving the way for physics-informed AI that can reliably assist in the design, analysis, and real-time monitoring of next-generation gas turbines.

\section{Conclusion}\label{sec8}
\noindent This survey provides a comprehensive evaluation of PINNs within the context of aerodynamics and aeromechanics, with a particular focus on their role in IGTs. The paper begins by identifying the limitations of conventional DL methods in scientific modeling and reviews how PINNs address these limitations by embedding physical laws such as governing equations and boundary conditions into deep learning architectures. Compared to conventional data-driven methods, PINNs have demonstrated a lower MSE and require significantly less data, often less than $50\%$ in physics-constrained environments.\medskip

\noindent The applications of PINNs in aerodynamic modeling are then examined, where they prove effective in simulating turbulent flows, shock interactions, and wake dynamics. The benchmarks show that PINNs can reconstruct flow fields with high fidelity, achieving fewer errors compared to high-fidelity DNS or RANS simulations. In the aeromechanical domain, PINNs have enabled accurate modeling of blade vibrations, flutter prediction, and fatigue crack propagation, often outperforming traditional methods in convergence speed and scalability, particularly for inverse problems or scenarios with limited or noisy measurements.\medskip

\noindent Recent advances, including adaptive sampling, dynamic loss weighting, and hybrid PINN architectures, have further improved the accuracy and computational efficiency of these models. In particular, adaptive sampling strategies have accelerated convergence, while hybrid methods have significantly reduced training times without compromising solution quality.\medskip

\noindent Despite these advances, several challenges remain. These include the lack of theoretical guarantees for convergence, difficulties in model error quantification, and computational inefficiencies in high-dimensional, multi-physics simulations. Future research should focus on robust error estimation techniques, neural network architecture optimization, and integration with multi-fidelity or domain decomposition methods to improve scalability and reliability.\medskip

\noindent Although PINNs may not fully replace traditional CFD or FEM solvers in all applications, they serve as a powerful complementary tool for real-time, data-limited, and inverse modeling tasks in IGTs systems. With ongoing methodological advances and improvements in computational efficiency, PINNs have the potential to significantly enhance modeling and simulation workflows in the aerodynamics and aeromechanics of next-generation IGTs.

\bibliographystyle{unsrt}  

\bibliography{references}

\end{document}